\documentclass[aps,prl,twocolumn]{revtex4-2}
\pdfoutput=1
\usepackage{graphicx}
\usepackage{grffile} 
\usepackage{dcolumn}
\usepackage{bm,amssymb,amsmath}
\usepackage[pdftex]{hyperref}
\hypersetup{colorlinks=true,citecolor=blue,linkcolor=red,urlcolor=blue}
\usepackage[all]{hypcap}
\usepackage{color}
\definecolor{red}{rgb}{0.8, 0.0, 0.0}
\definecolor{blue}{rgb}{0.06, 0.2, 0.65}
\definecolor{green}{rgb}{0,0.6,0}

\newcommand{\B}[1]{{\bm{#1}}}

\usepackage{float} 

\begin{document}
\title{Quasi-localized modes in crystalline high entropy alloys}
\author{Silvia Bonfanti$^1$, Roberto Guerra$^2$, Rene Alvarez-Donado$^1$, Pawel Sobkowicz$^1$, Stefano Zapperi$^{2,3}$, Mikko Alava$^{1,4}$}

\affiliation{
    $^1$ NOMATEN Centre of Excellence, National Center for Nuclear Research, ul. A. Soltana 7, 05-400 Swierk/Otwock, Poland
	\\$^2$Center for Complexity and Biosystems, Department of Physics, University of Milan, via Celoria 16, 20133 Milano, Italy
	\\$^3$ CNR - Consiglio Nazionale delle Ricerche,  Istituto di Chimica della Materia Condensata e di Tecnologie per l'Energia, Via R. Cozzi 53, 20125 Milano, Italy
	\\$^4$ Aalto University, Department of Applied Physics, PO Box 11000, 00076 Aalto, Espoo, Finland
}
\date{\today}

\begin{abstract}
High Entropy Alloys (HEAs) are designed by mixing multiple metallic species in nearly the same amount
to obtain crystalline or amorphous materials with exceptional mechanical properties. Here we 
use molecular dynamics simulations to investigate the role of positional and compositional disorder in determining the low-frequency vibrational properties of CrMnFeCoNi HEAs. Our results show that the expected dependence of the density of states on the frequency as $D(\omega)\sim\omega^4$ is recovered for amorphous HEAs and is
also observed for partially crystallized alloys with deviations that depend on the degree of crystallization. We find that the quasi-localized vibrations are still visible 
in crystalline HEAs, albeit suppressed compared to the corresponding amorphous alloys.
Our work offers a unified perspective to describe HEA mechanical properties in terms of their vibrational 
density of states.
\end{abstract}

\maketitle

\textit{Introduction. --} 
Metallic alloying has been the key factor in opening up new possibilities for designing materials with desired properties. The discovery of a new class of High Entropy Alloys (HEAs) represented a breakthrough in
the conventional strategy of alloy design: instead of having one or two main atomic species, as in traditional alloys, HEAs consist of several metallic species, five or more, in nearly equal concentration~\cite{george2019high}. Mixing many elements results in many
possible configurations that can be exploited in the search for those enabling
exceptional mechanical properties, e.g., improved hardness and a higher degree of fracture resistance~\cite{liu2022exceptional,otto2013influences}. HEAs are mostly studied as crystalline
materials, however, the research on their glassy disordered counterpart, namely HEA-glasses, is
also attracting a growing interest~\cite{wang2014high,chen2021high}. HEA-glasses display 
high strength and large resistance in high-temperature conditions.
In the literature, HEAs often refer only to crystalline configurations~\cite{yeh2004nanostructured,cantor2004microstructural} while their glassy counterparts are
denoted High Entropy Metallic Glasses \cite{chen2021high}. Here, we use the term HEAs to refer to the ensemble of the possible solid phases, from crystalline to partially crystallized up to amorphous ones.

\begin{figure}[tb!]
  \begin{center}
    \includegraphics[width=\columnwidth]{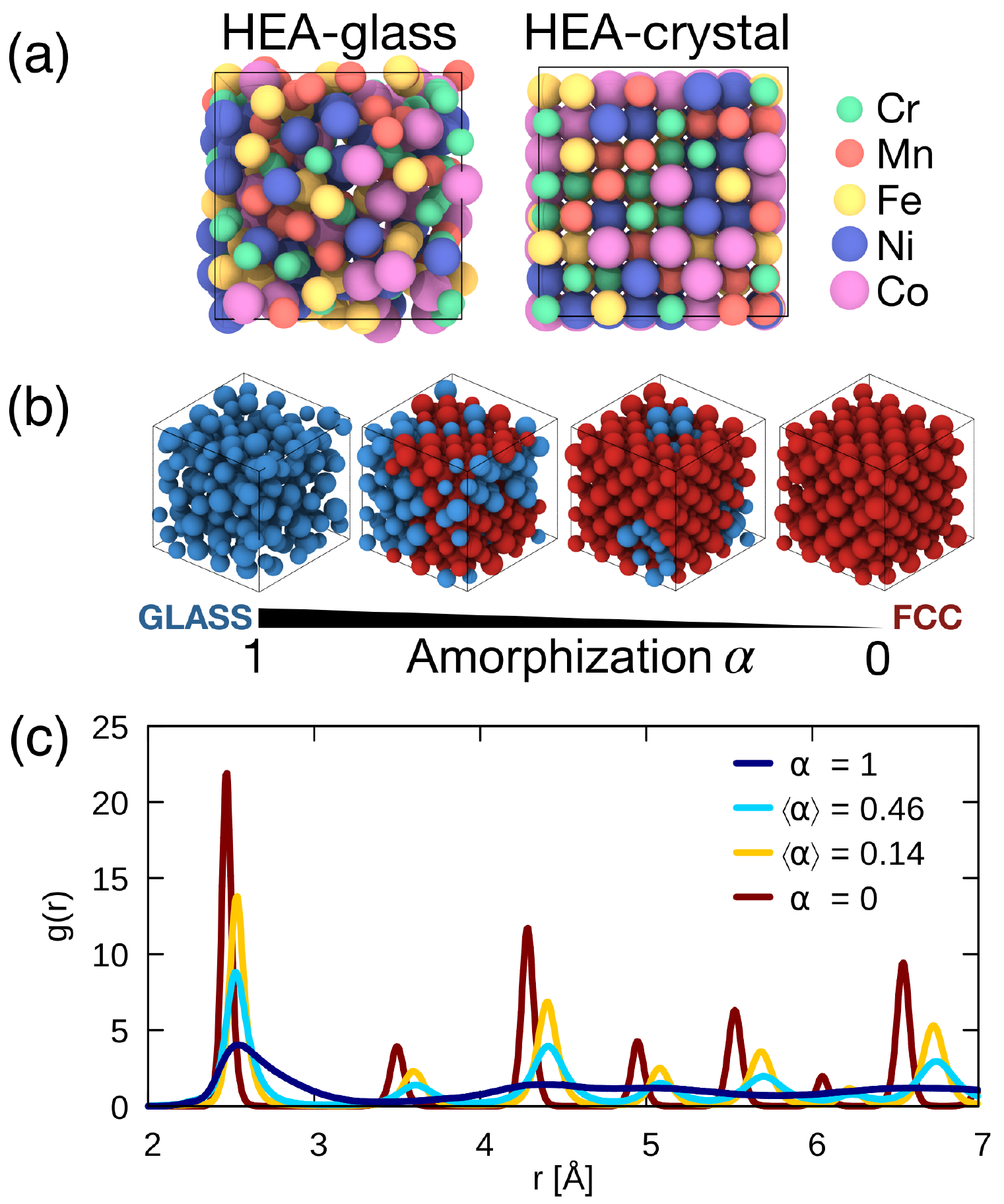}
  \end{center}
    \vspace{-6mm}
  \caption{(a) Concept of HEAs. (Left) HEA-glass, characterized by positional disorder. (Right) HEA–crystal characterized by positional order and compositional disorder: random distribution of the atom species on the lattice sites. Colors indicate different atomic species. (b)~Examples of configurations of Cantor HEAs obtained varying the level of positional disorder~$\alpha$: from glass (left) to intermediates (center) and crystal (right). The red color indicates the FCC structure (FCC), while the blue amorphous state (GLASS). (c)~Pair correlation function~$g(r)$ for different levels of amorphization.}
  \label{fig:order_disorder}
    \vspace{-3mm}
\end{figure}

From a structural perspective (see Fig.~\ref{fig:order_disorder}(a)), the distinctive feature of HEA-crystals is the presence of compositional disorder: atomic species are \textit{randomly} distributed across the lattice sites, avoiding repetitive patterns typical of traditional crystals. Amorphous HEAs are instead characterized by positional disorder \cite{binder2011glassy}.  The theoretical understanding of the structure-property relation of HEAs is currently missing, hampering the formulation of design principles to get desirable HEAs for target applications. 

In this Letter, we shed light on the vibrational properties of HEAs which are essential to rationalize the other physical properties.  There is ample evidence that the vibrational density of states $D(\omega)$ of glasses depends on the frequency~$\omega$ as $D(\omega) \sim \omega^4$ ~\cite{lerner2016statistics,17MSI,shimada2018spatial,wang2019low}, exceeding the Debye's contribution typical of three dimensional crystals that depends on $\omega^{2}$.
The associated low-frequency glassy modes are quasi-localized in space and are found to be related to the mechanical properties~\cite{richard2020predicting,richard2021simple} and the stability of glasses~\cite{rainone2020pinching}. 
The law is found to hold independently from the model of glass former, being valid for toy models of glasses with binary interactions~\cite{15BMPP,angelani2018probing,17MSI,18KBL,Moriel2019,lerner2021low} and for more realistic models of silica glasses \cite{bonfanti2020universal,20GRKPVBL}. 
The quartic law has been shown to be independent of dimensionality~\cite{wang2021low,wang2022scaling} and temperature, as long as the system remains within its inherent structure~\cite{guerra2022universal}.
The universal form of the quartic law is believed to be a consequence of the positional disorder as 
suggested by a recent numerical work~\cite{lerner_2022} where the level of positional disorder has been controlled in Lennard-Jones models of a binary alloy.
In this letter, we investigate the validity of the quartic law for HEAs and explore the role of compositional disorder by simulating the low frequency vibrational properties of the Cantor alloy~\cite{cantor2004microstructural}, composed by (Cr,Mn,Fe,Co,Ni) in equiatomic proportion,
for different degrees of positional disorder.

\textit{System and protocol. --} Simulations are performed in three dimensions with periodic boundary conditions. The interaction between the atoms is given by the Modified Embedded Atom Method (MEAM) interatomic potential, implemented in LAMMPS~\cite{lammps}. MEAM is an extension of the class of Embedded Atom Method (EAM) potentials, developed to describe atomic interaction in metallic systems~\cite{daw1984}. The potential is constructed to have continuous second derivatives to avoid spurious cutoff effects in the Hessian calculation. 
For Cantor HEAs, we use the parameters of Ref.~\cite{choi2018}. 
Units are defined based on energy, length, and time in eV,~\AA, and ps. The choice of the system size to be sufficiently small is made with the aim to disentangle quasi-localized modes from phonons as stressed in previous works~\cite{bonfanti2020universal,lerner2021low}. 
\textit{HEA-Glasses:} We generate a set of 9000 samples, each with $N$=200 atoms randomly placed with a density of 6.65$\times$10$^{-24}$\,g/\AA$^3$~\cite{karimi2022}. After an initial 2~ps of damped Newtonian dynamics with Lennard-Jones interatomic interaction, we switch to the potential for Cantor HEAs and perform subsequent 8~ps of damped Newtonian dynamics, and then equilibrate the system at 2000~K. Finally, we quench to 0~K by structurally relaxing the samples using the fast inertial relaxation engine (FIRE)~\cite{bitzek2006structural} minimization until the maximum force on each atom is smaller than $10^{-10}$\,eV/\AA.
\textit{HEA-Crystals:} Since Cantor HEA-crystals possess face-centered cubic (FCC) symmetry, we start from a pure FCC lattice of Ni atoms with $N=256$, and we randomly change the atomic species until we obtain equiatomicity, i.e., an equal fraction of elements. 
We run an initial Lennard-Jones dynamics and then switch to the potential for Cantor HEAs. We then equilibrate the system at 500~K and instantaneously quench to 0~K with FIRE and the above constraint on the forces as stopping criteria. The dataset consists of 920000 samples.
\textit{HEA-Intermediates:} To model partially crystallized HEA configurations, we vary the level of positional disorder through a variant of the procedure introduced in Ref.~\cite{goodrich2014solids}. Starting from one FCC configuration of HEA-crystal with $N=256$ particles, we randomly remove a selected number~$m$ of particles and re-insert them in a random position in the box. In our simulations, we choose $m=1,5,10$ and produce respectively about 447000, 256000, and 41000 samples for each case. Subsequently, we relax the size of the simulation box to adjust to the new volume and perform again FIRE~\cite{bitzek2006structural} minimization with the force constraint to relax the system. We further define a parameter $\alpha=1-\frac{N_{FCC}}{N}$, where $N_{FCC}$ is the number of atoms with FCC symmetry, as resulting from adaptive common neighbor analysis calculation~\cite{stukowski2012structure} using a cutoff radius of 4\,\AA, which assures the embracement of second neighbors. We checked that all the glass (crystal) structures give $\alpha=1$ ($\alpha=0$). The intermediate configurations with a given $m$ produce, after the structural relaxation, a broad range of $\alpha$ values, as shown in Supplementary Figure~S1. Therefore, we restrict the considered amorphization values by selecting two ranges of $\alpha$, 0.1--0.3, and 0.4--0.7, corresponding to two sets of samples with average $\langle\alpha\rangle=0.14$ and $\langle\alpha\rangle=0.46$, respectively.
Detail of the distribution of the number of FCC structures 
for different values of~$m$ is reported for completeness in Supplemental Material Fig.~S1. 

\textit{Vibrational modes calculation. --} The potential energy of the system $U(\B r_1, \cdots \B r_N)$, with $r_i$ being the i$th$ coordinate of one atom, provides the Hessian matrix~$\B H$, which gives, in the harmonic approximation, the frequencies and associated eigenmodes~\cite{06ML}
\begin{equation}
	H_{ij}^{\alpha \beta} \equiv \frac{1}{\sqrt{m_i m_j}} \frac{\partial^2 U(\B r_1, \cdots \B r_N) }{\partial r_i^\alpha \partial r_j^\beta} \ .
	\label{Hessian}
\end{equation}
where $m_i$ is the mass of the $i$th atom. By diagonalizing~$\B H$, we obtain the modes and eigenvalues in athermal conditions. The frequencies~$\omega_i$ are obtained from the square root of the eigenvalues.

\textit{Results. --} Figure~\ref{fig:order_disorder}(b) reports examples of snapshots of Cantor HEAs for different values of the amorphization parameter $\alpha$: from left to right, HEA-glass with $\alpha=1$, two HEA-intermediates with $\langle\alpha\rangle=0.46,0.16$ and HEA-crystal with $\alpha=0$. Note that the different atomic species are marked with different sizes of radius.
The color represents the number of FCC structures of the configuration: red atoms possess FCC local positional order while blue atoms disorder.
The evolution of the pair correlation function $g(r)$ for different levels of amorphization is reported in Fig.~\ref{fig:order_disorder}(c). 
Note that the pair correlation function has been averaged over 8500 samples. We observe that the shape of the $g(r)$ is consistent with a glass (blue curve) and reaches the typical FCC pattern for the crystal (red curve) as the amorphization is reduced (respectively light blue and yellow curves for $\langle\alpha\rangle$=0.46,0.14).

\begin{figure}[tb!]
  \begin{center}
    \includegraphics[width=\columnwidth]{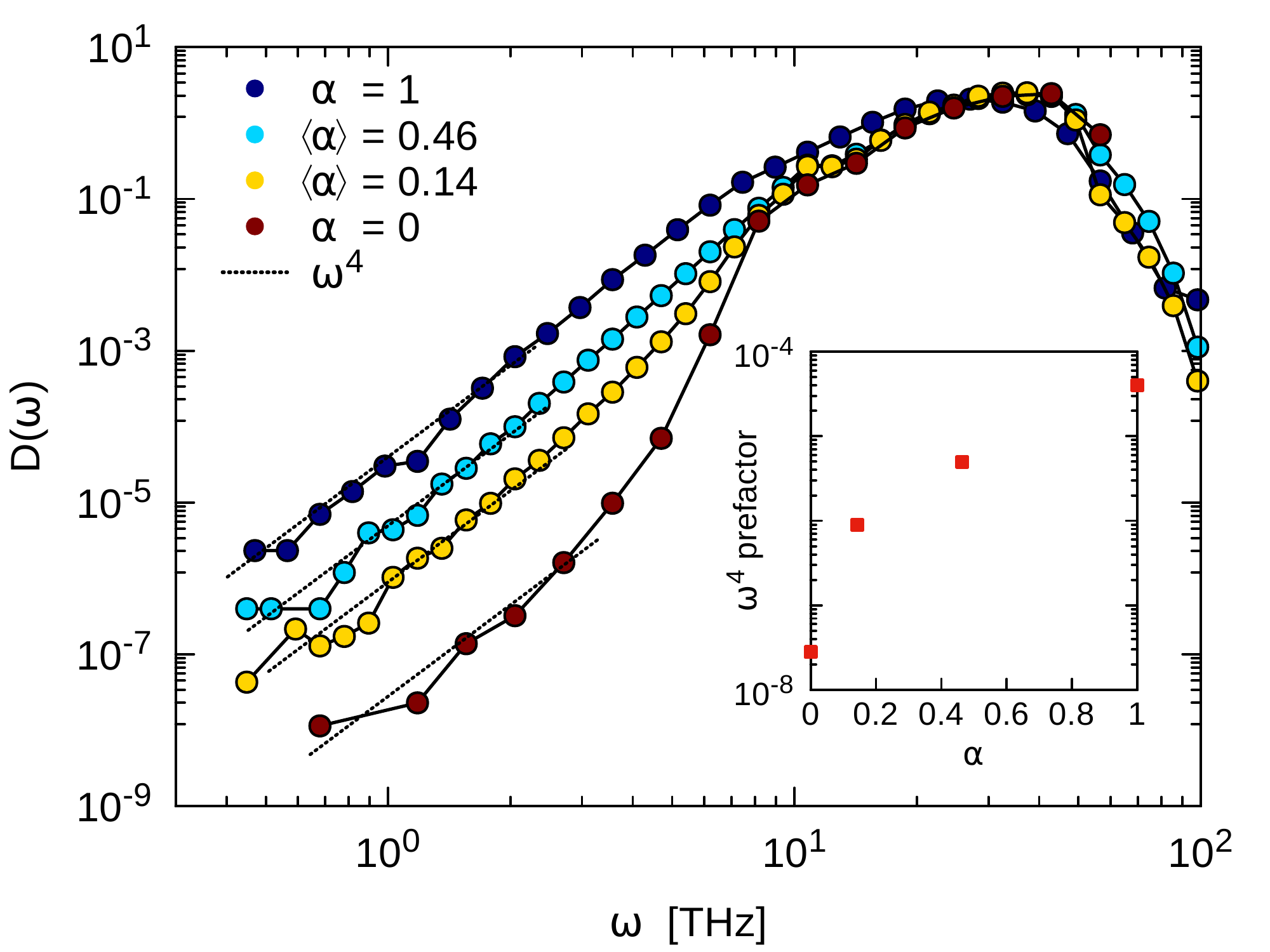}
  \end{center}
  \vspace{-6mm}
  \caption{Vibrational density of states for HEAs with different amorphization degree $\alpha$ ($\alpha$=0~crystal, $\langle\alpha\rangle$= 0.46, 0.14~intermediates, $\alpha$=1~glass). The inset shows the fitted values of the prefactor of the density of states versus $\alpha$.}
  \vspace{-3mm}
  \label{fig:figure_dos}
\end{figure}

The density of states for different levels of disorder: from glass to crystal, is shown in Fig.~\ref{fig:figure_dos}, which is the main result of this work. 
Amorphous samples ($\alpha=1$, blue dots) display the $\omega^4$ law 
typical of glassy systems, confirming that also HEA-glasses show the same universality class as other glassy systems. 
By decreasing the level of disorder, thus reducing the level of amorphization ($\langle\alpha\rangle$=0.46,0.16, light blue and yellow curves respectively), we find that also the partially crystallized configurations show $\omega^4$ trend. However, the magnitude of the density of states also decreases with decreasing $\alpha$.  
The suppression is even more evident in the case of HEA-crystals ($\alpha=0$, red dots), where the $\omega^4$ law is only 
present at very low frequencies. 

The inset of Fig.~\ref{fig:figure_dos} displays the prefactor of the density of states for HEAs as a function of the amorphization on a log-lin scale, obtained from the fits of the curves with $\omega^4$ at low frequency.  The prefactor of HEA-crystals interestingly shows a drop that deviates from the amorphous and partially crystallized cases.
 We checked, however, that choosing slightly different range values for $\alpha$ or changing the number of samples for the average does not significantly affect the results. Note also that the frequency range has been binned in a logarithmic way and changing the binning to linear or the size of the binning does not alter the results. 

To further examine the scaling range of the $\omega^4$ law, we consider the average of the minimal frequency~$\omega_{\rm min}$ for each configuration over the ensemble of configurations is indicated as $\langle\omega_{\rm min}\rangle$.
Extreme value theory implies that the distribution of $\omega_{\rm min}$ should follow the Weibull distribution~\cite{karmakar2010statistical},
\begin{equation}
	W(k;\omega_{\rm min})=\frac{(k+1)(\Gamma(1.2))^{k+1}}{\langle \omega_{\rm min}\rangle^{k+1}}~\omega_{\rm min}^{k}~e^{-\left(\frac{\omega_{\rm min}\Gamma(1.2)}{\langle \omega_{\rm min}\rangle}\right)^{k+1}} \ ,
	\label{Weib}
\end{equation}
where $\Gamma(x)$ is the Gamma function, $\Gamma(1.2)\approx 0.918$.

\begin{figure}[tb!]
  \begin{center}
    \includegraphics[width=\columnwidth]{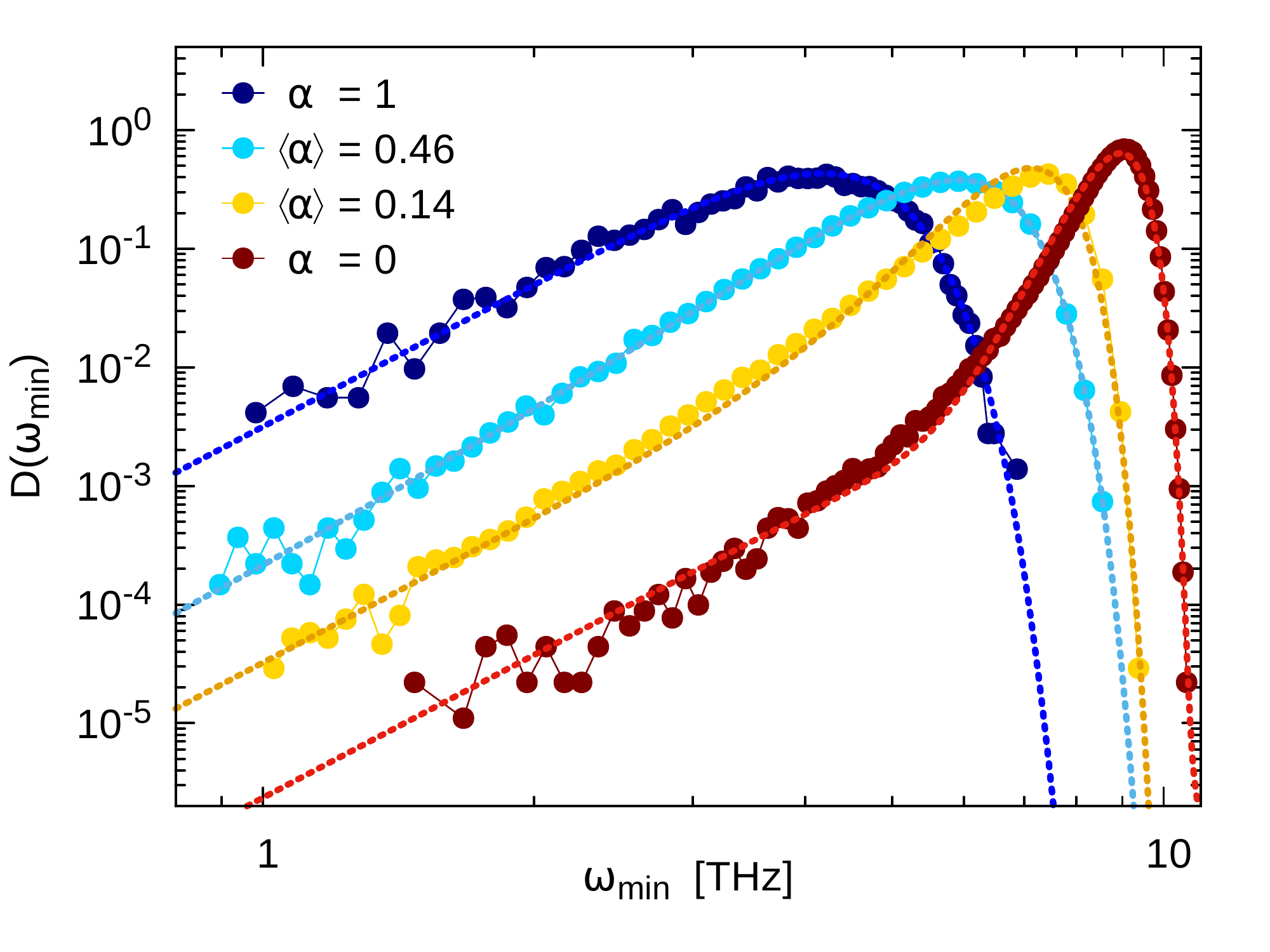}
  \end{center}
  \vspace{-6mm}
  \caption{Distribution of the minimum eigenfrequency $\omega_{\rm min}$ for different amorphization values~$\alpha$. The dotted lines are fit via a double Weibull distribution. Fit and data were normalized to unity.}
  \vspace{-3mm}
  \label{scaleplot}
\end{figure}
For amorphous materials, the parameter $k$ glasses should assume the well-defined value of $k=4$, in agreement with the power-4 scaling law holding therein. The value of $k$ is expected to increase in poorly disordered materials, corresponding to a faster crack propagation and narrower $\omega_{\rm min}$ distributions. In the limit of an ideal crystal, one would obtain a unique $\omega_{\rm min}$ value, with $W(\omega_{\rm min})$ approaching the $\delta$ function (i.e.\ $k\to\infty$).
In our case of partially crystallized HEAs, we therefore expect that a combination of (at least) two $W$ distributions is required to account for the multiple failure rates associated with the different constituent phases.

This prediction is tested in Fig.~\ref{scaleplot}, where the fitting of the data is performed by two Weibull functions, one with fixed $k=4$, and another with free $k$ parameter (See Supplementary Fig.~S2 for a detailed plot of the fitting functions).
We note that for decreasing disorder ($\alpha \to 0$), the peak of the distribution shifts toward higher frequencies, consistently with the appearance of delocalized high energy modes. By calculating the participation ratio~$P$~\cite{bonfanti2019elementary} of the eigenvectors associated to the $\omega_{\rm min}$ for different $\alpha$ we have obtained an average $\langle P\rangle = 0.05, 0.13, 0.18, 0.37$ for $\alpha = 1, 0.46, 0.14, 0$, respectively. The distributions $D(P)$ (see Supplementary Fig.~S3) are peaked at around $P_l\sim 0.05$ (strong localized modes) for larger amorphous fraction, while a second peak at $P_d\sim 0.45$ (weak localized modes) starts to appear when approaching the crystalline phase, becoming dominant only at $\alpha=0$. Remarkably, while the $D(P)$ associated with $\alpha=1$ shows no trace of contributions from $P_d$, the one associated with $\alpha=0$, clearly shows some contribution from $P_l$ which we address to the residual atomic positional perturbation induced by the compositional disorder.

\begin{figure}[t!]
  \begin{center}
    \includegraphics[width=\columnwidth]{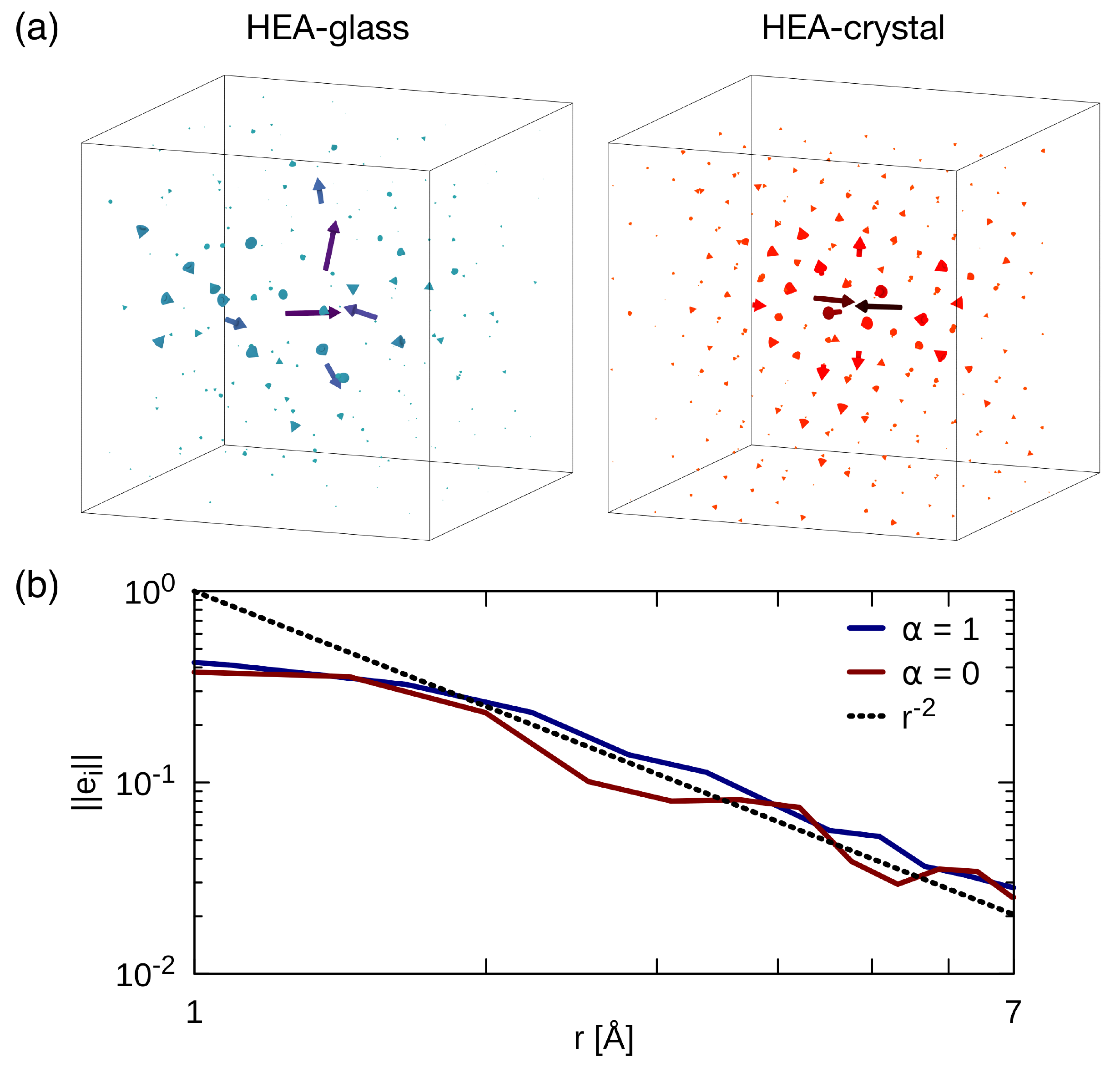}
      \vspace{-6mm}
       \end{center}
  \label{fig:event}
  \caption{(a) Orthogonal view of the eigenvectors~$\hat{e}$ corresponding to the $\omega_{\rm min}$ for typical configurations of HEA-glass (left) and HEA-crystal (right). Arrows size has been increased by a factor of 4. The color of the arrows represents the modulus of the vectors: from dark blue ($e=0.5$) to light blue $e=0$ for HEA-glasses and from dark red ($e=0.5$) to light red $e=0$) for HEA-crystals. (b) Spatial decay of the quasi-localized modes for HEA-glasses and HEA-crystals averaged over ten configurations.}
    \vspace{-3mm}
\end{figure}

Examples of quasi-localized modes for amorphous and crystalline HEAs are shown in Fig.~\ref{fig:event}(a) (left and right panels, respectively). The arrows indicate the eigenvectors corresponding to the smallest frequency~$\omega_{\rm min}$. Specifically, $\omega_{\rm min}$=0.9974~THz and the participation ratio $P$=0.0415 for the glass and $\omega_{\rm min}$=1.4912~THz and $P$=0.0522 for the crystal system. 
Arrows are colored according to the magnitude of the modulus of the vectors. 

After visualizing the quasi-localized modes~$\hat{e}$ in real space, we focus on the analysis of their spatial decay for both HEA-glasses and -crystals. We calculate the core of the event by computing the weighted average of the position of the modes with their magnitude, considering only the six particles with the highest $||\hat{e}_i||^2$. 
The spatial decay is defined as the distance~$r_i$ of each particle from the core. We consider the average over ten different configurations and bin the value of the distance. 
Quasi-localized modes are expected to follow $r^{-2}$ at the far field~\cite{lerner2021low}, and our results confirm the same trend for both the HEA-glasses and HEA-crystals, see Fig.~\ref{fig:event}(b).

Finally, we also computed the participation ratio of all the modes and found that quasi-localized low frequency modes are characterized by a low participation ratio. This result agrees with previous investigation on glassy systems~\cite{lerner2021low} (see also Supplementary Fig.~S3).

\textit{Summary and Conclusions. --} The goal of this work is to examine the low frequency vibrational properties of HEAs for different levels of amorphization
and study to what extent the $\omega^4$ law remains valid. Our results show that the density of states at low frequency follows the $\omega^4$ non-phononic trend associated with quasi-localized modes in space for all levels of positional disorder. We find that compositional disorder is the leading cause of the presence of quasi-localized modes and observe the $\omega^4$ law also in crystalline HEAs. This is a remarkable distinction from the density of states of conventional crystals that follows the $\omega^2$ trend in three dimensions. 
HEA-crystals can therefore be regarded as intermediate between completely disordered solids (specifically metallic glasses) and ordered solids (pure crystalline metals). 
Experimentally, glassy-like behaviors have been reported in other types of disordered crystals characterized by a small amount of disorder at low temperatures, e.g.~\cite{moratalla2019emergence}. Literature on HEA-crystals at low temperatures is not extensive~\cite{wang2019incredible,naeem2021temperature}, but there might be a possibility that those glassy anomalies are present there since we find here 
the presence of quasi-localized modes.\\
The compositional disorder in HEA-crystals implies that atoms are slightly displaced from the ideal lattice sites causing lattice distortions \cite{zhang2014microstructures} as also revealed experimentally from single-crystal synchrotron X-ray diffraction for the equiatomic Cantor HEA-crystals~\cite{okamoto2016atomic}. The measured value for the squared atomic displacement parameter from the ideal FCC sites of a pure metal and at low temperature is 23.5$\pm$0.4~pm${^2}$~\cite{okamoto2016atomic}.    
Quasi-localized modes, however, reveal to be a tool to measure the local disorder being able to capture even the tiny lattice distortions present in HEA-crystals. The number of localized modes decreases with increasing order. 
A separate discussion is necessary for the relation between the vibrational properties of HEA-crystals and lattice distortion, which is a task for future research. \\
{\it Acknowledgements:} We gratefully thank Prof. Itamar Procaccia for fruitful discussions. SB, RAD, PS and MJA are supported by the European Union Horizon 2020 research and innovation program under grant agreement no. 857470 and from the European Regional Development Fund via the Foundation for Polish Science International Research Agenda PLUS program grant No. MAB PLUS/2018/8.

\bibliographystyle{unsrt}
\bibliography{omega4hea_supp}

\clearpage

\pagebreak
\newpage
\widetext
\begin{center}
\textbf{\large Supplementary Material: Quasi-localized modes in crystalline high entropy alloys}
\end{center}
\setcounter{equation}{0}
\setcounter{figure}{0}
\setcounter{table}{0}
\setcounter{page}{1}
\makeatletter
\renewcommand{\theequation}{S\arabic{equation}}
\renewcommand{\thefigure}{S\arabic{figure}}
\renewcommand{\bibnumfmt}[1]{[S#1]}
\renewcommand{\citenumfont}[1]{S#1}
\begin{figure}[h]
  \begin{center}
    \includegraphics[width=0.5\textwidth]{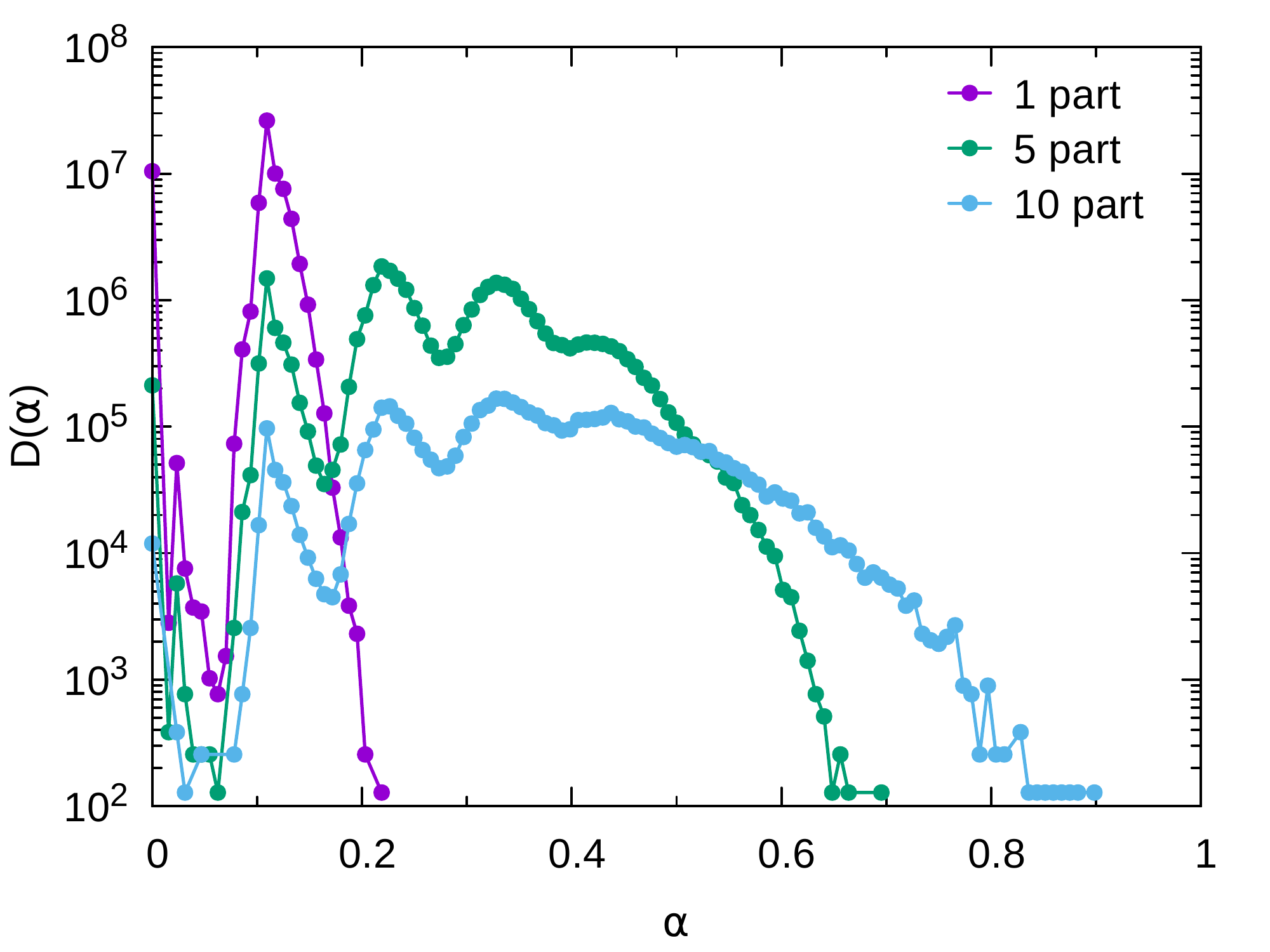}
  \end{center}
  \caption{Distribution of the amorphization degree $\alpha$ for the structures obtained by randomly displacing $m = $ 1, 5, or 10 particles starting from an FCC HEA crystal.}
  \label{fig.figS1}
\end{figure}
\begin{figure}[h]
  \begin{center}
    \includegraphics[width=0.5\textwidth]{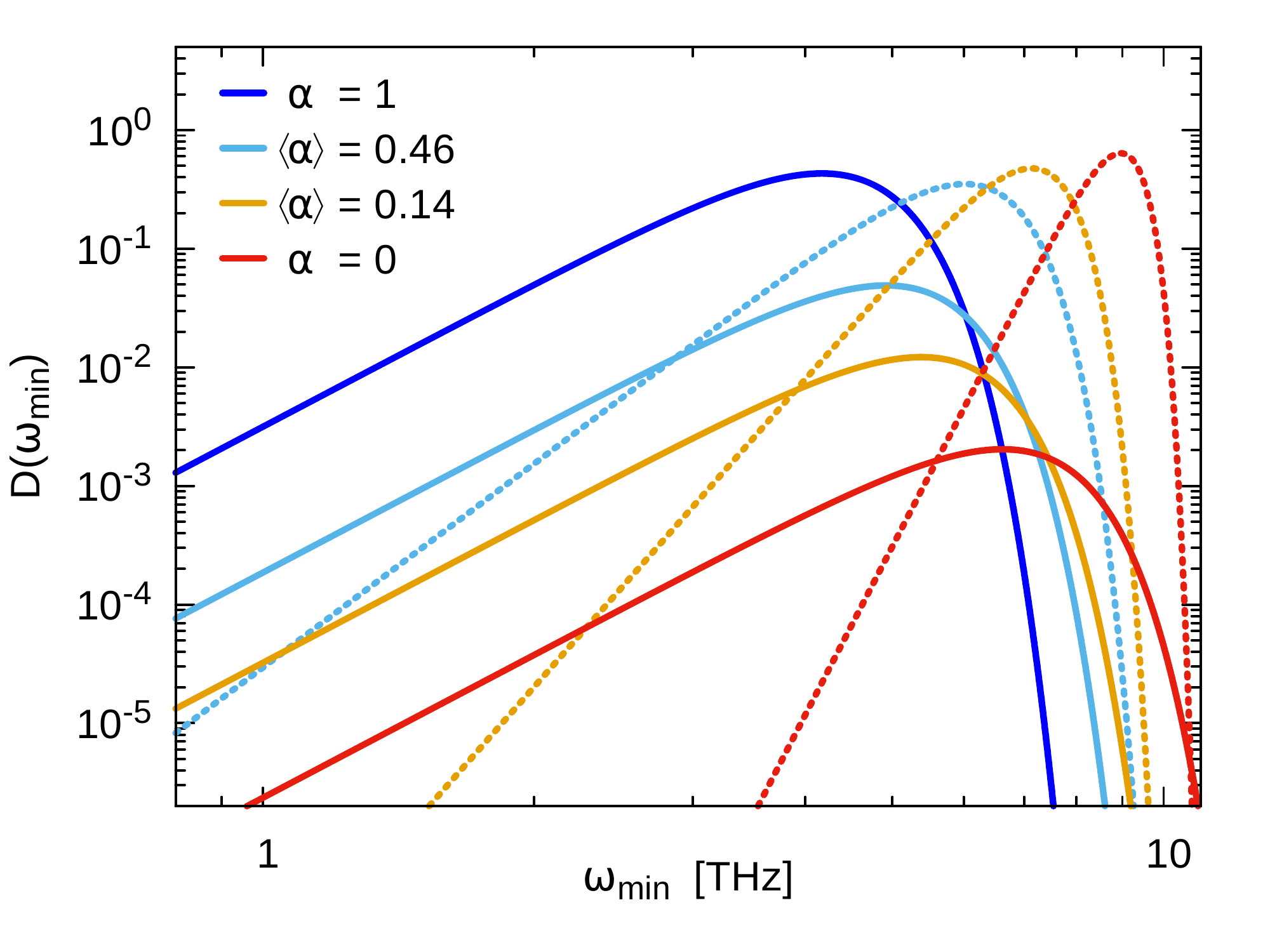}
  \end{center}
  \caption{
    For each set with amorphization degree $\alpha$ are reported the two Weibull functions whose sum gives the fit of Fig.~3. In the case of pure glass ($\alpha=1$) the two Weibull functions result degenerate.
  }
  \label{fig.figS2}
\end{figure}
\begin{figure}[h]
  \begin{center}
    \includegraphics[width=0.5\textwidth]{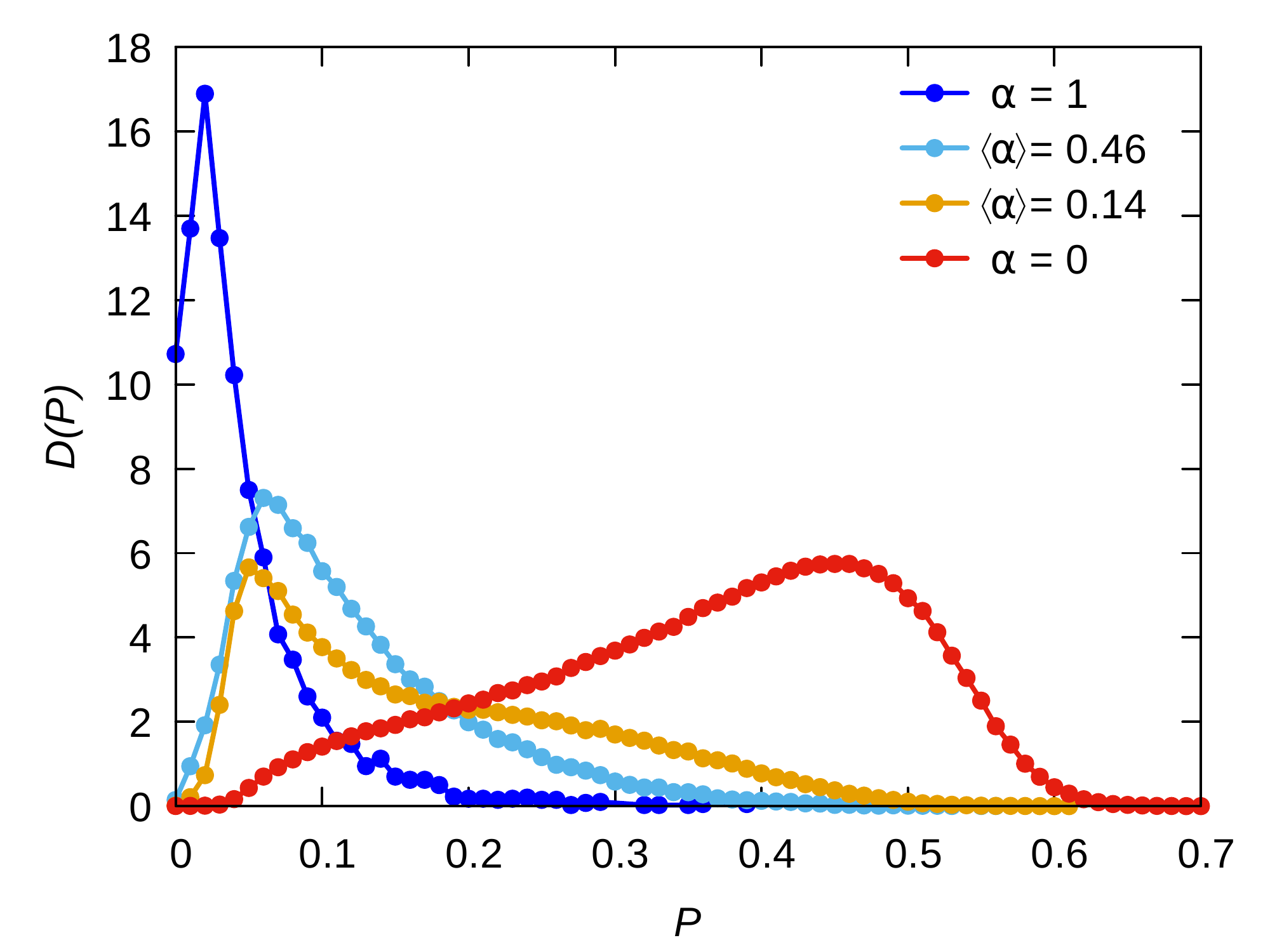}
  \end{center}
  \caption{Distribution of the participation ratio for each set of samples with different amorphization degree $\alpha$. }
  \label{fig.figS3}
\end{figure}

\end{document}